\documentstyle[amsfonts,aps,twocolumn,epsf]{revtex}

\newcommand{\beq}{\begin{equation}}
\newcommand{\eeq}{\end{equation}}
\newcommand{\beqa}{\begin{eqnarray}}
\newcommand{\eeqa}{\end{eqnarray}}
\newcommand{\non}{\nonumber}

\begin{document}

\title{Analytic computation of the Instantaneous 
Normal Modes spectrum in low density liquids}

\author{
Andrea Cavagna,
\thanks{E-mail: a.cavagna1@physics.ox.ac.uk}
Irene Giardina
\thanks{E-mail: i.giardina1@physics.ox.ac.uk} }
\address{Theoretical Physics, University of Oxford,
1 Keble Road, Oxford, OX1 3NP, United Kingdom}
\author{
Giorgio Parisi
\thanks{E-mail: giorgio.parisi@roma1.infn.it}
}
\address{Dipartimento di Fisica, Universit\`a di Roma La Sapienza and
INFN Sezione di Roma I, P.le Aldo Moro 5, 00185 Roma, Italy}

\date{\today}

\maketitle

\begin{abstract} {\bf Abstract:}
{\sl We analytically compute the spectrum of the Hessian of the Hamiltonian
for a system of $N$ particles interacting via a purely repulsive
potential in one dimension. Our approach is valid in the low density
regime, where we compute the exact spectrum also in the localized sector.
We finally perform a numerical analysis of the localization properties of
the eigenfunctions.}
\end{abstract}

\vskip 0.3 truecm

Great efforts have recently been done to find a comprehensive 
microscopic theory able to describe the behaviour of super-cooled 
liquids \cite{angell}. 
Among the others, a key point is to understand the mechanism of the 
glass transition, and to find a description suitable both for the 
super-cooled liquid and for the amorphous solid which forms at low 
temperatures.

The general frame where a great number of recent analysis have been
performed is the Instantaneous Normal Modes (INM) approach \cite{keyes}.
The main idea of this approach is that liquids are `solid-like' at
short times $t<\tau$, where the typical diffusion time $\tau$ increases 
strongly when lowering the temperature. 
Liquids' dynamics would correspond 
in this picture to vibrations about some equilibrium 
positions with periodic jumps into new local minima\cite{inm}.
In order to describe in a quantitative way this behaviour,
it is important to study the properties of the Hessian matrix of 
the Hamiltonian, averaged over the equilibrium 
Boltzmann distribution. 
The crucial quantity is the typical {\it spectrum} 
of the Hessian, whose eigenvectors are the so-called 
Instantaneous Normal Modes. The determination of the spectrum 
therefore represents a vital task for any theoretical study 
of liquids.

The aim of this letter is to outline an analytic approach for the  
computation of the INM spectrum.
Some important steps in this direction have been recently done in 
\cite{wuloring} and \cite{strattI}, where a remarkably good 
agreement with numerical simulations on liquids has been achieved
and where more realistic models than the one we study here were 
considered. 
However, these former analytic computations of the INM spectrum 
assumed a {\it Gaussian} probability distribution for the auxiliary 
degrees of freedom of the liquid \cite{strattI} (this technical point 
will be clarified later) and we think that a Gaussian approach is likely 
to be too simple for a study of non-trivial spectral properties, such 
as localization.
This is a crucial difference with the computation we present
here, which, going beyond the Gaussian approximation, allows us to 
better investigate the spectrum in the localized sector.

Let us consider a system of $N$ interacting particles with Hamiltonian
$H=\sum_{k>l}^N V(r_{kl})$,
where $V(r)$ is a two-body potential. 
The Hessian matrix $\bf A$ is defined by,
$A_{kl}^{\mu\nu} = \partial_k^\mu \partial_l^\nu H$,
with $\mu, \nu=1,\dots,d$, being $d$ the dimension of the space.
In the present letter we shall consider the one dimensional case $d=1$, 
in order to keep the algebra as simple as possible. 
The form of $\bf A$ is
$A_{kl} = -J_{kl}+\delta_{kl}\sum_i^N J_{ki}$, where $J_{kl}= V''(r_{kl})$.
The diagonal term of $\bf A$ is a consequence of the translational 
invariance of the system, which requires $\sum_k^N A_{kl}=0$. 

As a first step in our analysis, we must 
find out what is the probability 
distribution of the matrix $\bf J$ induced by the equilibrium 
probability over the positions of the particles. 
We introduce here our first approximation, 
assuming that the probability distribution $P[{\bf J}]$ is factorized into 
the single probabilities of the particles pairs. In this way the elements
of ${\bf J}$ (but not of $\bf A$) are independently distributed, i.e.
$P[{\bf J}]\equiv\prod_{k>l}^N p(J_{kl})$.
This approximation works well at low densities, where the three-particles
correlations are negligible.

Once assumed this factorized form for $P[{\bf J}]$, we can express 
the pair-probability $p(J)$ as,
\beq
p(J) = \frac{\rho}{N} \int dr \;  g^{(2)}(r) \; \delta(J-V''(r)) \ ,
	\label{uno}
\eeq
where $g^{(2)}(r)$ is the two-particles correlation function and
$\rho$ is the density. In the following we shall consider the 
soft-spheres potential $V(r)=1/r^m$. In this case density and
temperature scale homogeneously, so that we can simply fix 
the density $\rho=1$ and consider the high temperature regime
as equivalent to the low density one. This does not hold for a 
non-homogeneous potential and what follows must be interpreted 
as a low density calculation.
 
In the definition of $p(J)$ we must insert the explicit form of 
$g^{(2)}(r)$, which can be obtained by means of some approximation 
schemes in liquid theory or by numerical simulations. 
In the present context we want to show what are the results of our method 
in the simplest analytical way, so that we stop at the first order of the
virial expansion and assume $g^{(2)}(r)=\exp(-\beta V(r))$. 
Thus, from eq.(\ref{uno}) we have, 
\beq
p(J) \sim \frac{1}{N}\; \frac{e^{-\beta_s J^{b}}}{J^{1+c}}\equiv
	\frac{1}{N} \; q(J) \ ,
	\label{qj}
\eeq
with $\beta_s=\beta \; [m(m+1)]^\frac{1}{m+2}$, $b=m/(m+2)$
and $c=1/(m+2)$. For realistic values of $m$ (typically $m=12$) the
parameter $b$ is very close to one. Therefore, we will directly set
$b=1$ in $p(J)$ in order to simplify our calculation. 
We will show in the discussion of our results that the actual 
spectrum is very weakly dependent on this approximation.
As it stands the distribution $p(J)$ is not normalizable, but we can
regularize it in the following way. Let us put an IR cut-off $\bar r$, 
by setting $V(r)=0$ for $r>\bar r$, and let $\eta=V(\bar r)$. We obtain in
this way a regularized form of the pair probability:
\beq
p_\eta(J) = \delta(J) + \frac{1}{N} 
\left( q(J)\; \theta(J-\eta) - 
\delta(J) \int_\eta^\infty dJ' \; q(J') \right) 
	\label{pj}
\eeq
where $q(J)$ is defined in equation (\ref{qj}).
In the following we shall use the notation,
\beq
\langle f(J) \rangle \equiv \int_\eta^\infty dJ \; q(J) \; [f(J)-f(0)]
	\label{lupu}
\eeq
Note an important point: the distribution  $p_\eta(J)$ is 
{\it diluted}, since the probability of finding an element 
of the matrix $\bf J$ larger than $\eta$ is of order $1/N$.

In order to compute the spectrum of $\bf A$ we introduce 
the resolvent, 
$G_{kl}(\lambda|{\bf J}) = [\lambda - {\bf A} +i\epsilon]_{kl}^{-1}$.
In the following we shall include the small imaginary term $i\epsilon$ 
in $\lambda$.
The spectrum (or density of states) $D(\lambda)$ is then given by,
$D(\lambda)=\lim_{\epsilon\to 0}{\rm Im}\;{\rm Tr}\;{{\bf
G}}(\lambda)/\pi N$,
where ${\bf G}(\lambda)$ is the average over ${\bf J}$ of the 
resolvent. Following the recursive method of \cite{cizeau}, we
write now a self-consistent equation for $\bf G$.
Given a system of $N$ particles, we add an extra particle, with label
$0$, and we write the expression of the $(0,0)$ component of $\bf G$ in
the $(N+1)$-particles system, isolating the contribution of the $0$-th 
particle:
\[
iG^{(N+1)}_{00}(\lambda|{\bf J}) 
= \frac{1}{Z}\int d\phi \; d\psi_1\dots d\psi_N \; \phi^2 
	\; e^{S(\phi,\psi_1\dots \psi_N)} \ \ \ , \ \ \
	\non
\]
where $Z$ is the effective partition function and where $\phi\equiv\psi_0$.
The action $S$ is given by,
\[
S=	\frac{i}{2}\lambda\phi^2
	-\frac{i}{2}\sum_{k=1}^N J_{0k}(\phi-\psi_k)^2
	+\frac{i}{2}\sum_{k,l=1}^N \psi_k [G^{(N)}]^{-1}_{kl} \psi_l \ .
	\non
\]
In the same spirit as the cavity method \cite{spin}, 
we have written a self-consistent equation which 
relates the element $G_{00}^{(N+1)}$ 
of the system with $(N+1)$ particles, to the matrix $G_{kl}^{(N)}$ of the
system with $N$ particles.
We note that $G_{kl}^{(N)}$ is the resolvent of the system in
absence of the $0$ particle and therefore it 
does not depend on the new vector $J_{0k}$.
Besides, it is shown in \cite{cizeau} that only the diagonal 
elements contribute to the last sum in the action $S$.

Since we have to perform an average over ${\bf J}$, 
it is useful to introduce replicas, by writing 
$Z^{-1}=\lim_{n\to 0} Z^{n-1}$. After replication we have:
\[
iG_{00}^{(N+1)}(\lambda |J_{0k},G_{kk}^{(N)})=
\lim_{n\to 0} \
\frac{1}{n} \int d\vec\phi \ (\vec\phi)^2 \
\Omega(\vec\phi|J_{0k},G_{kk}^{(N)})   \ ,
\]
where $\vec\phi\equiv(\phi_1\dots\phi_n)$ and
\beqa
&&\phantom{zibidopela}
\Omega(\vec\phi|J_{0k},G_{kk}^{(N)})=
e^{i\frac{\lambda}{2}\sum_a^n\phi_a^2} \times
\non \\
&&\int \prod_{k=1}^N d\vec\psi_k \ 
e^{-\frac{i}{2}\sum_{k,a}^{N,n}J_{0k}(\phi_a-\psi_{k,a})^2} 
e^{\frac{i}{2}\sum_{k,a}^{N,n}\psi_{k,a}^2 [G^{(N)}]^{-1}_{kk}} \ .
	\label{azu}
\eeqa
The crucial quantity is now the average probability distribution 
$\Omega(\vec\phi)$, obtained by averaging expression (\ref{azu}) over 
$\bf J$.
Indeed, from $\Omega$ we can reconstruct the average resolvent ${\bf G}$ 
by integration over $\vec\phi$ and therefore obtain the spectrum.
Thus, unlike what has been done in \cite{cizeau}, we shift our 
attention on the pursue of a self-consistent equation for $\Omega$ and 
not for the resolvent $\bf G$ itself. Let us note that it is this very 
distribution $\Omega(\vec\phi)$ that, 
as said in the introduction,
has been assumed to be Gaussian in the calculations of \cite{wuloring} 
and \cite{strattI} 
(the distribution $s(\rm X)$ in Section III.B of \cite{strattI} is exactly
the same object as our $\Omega(\vec\phi)$).

The dependence on the matrix 
${\bf J}$ of $\Omega$ is divided in two independent parts: 
the vector $J_{0k}$ and the matrix ${\bf G}^{(N)}$. 
We can therefore compute separately the two averages and, after some
algebra, we have:
\[ 
\Omega(\vec\phi)=
e^{i\frac{\lambda}{2}\sum_a^n\phi_a^2}
\exp\left(\int d\vec\psi \; \Omega(\vec\psi) \langle 
e^{-\frac{i}{2}J\sum_{a}^{n}(\phi_a-\psi_{a})^2} \rangle\right) \ .
\]
We can now obtain a more tractable integral equation by defining the 
following function $g$:
\[
g(\vec\phi)\equiv\int d\vec\psi \; \Omega(\vec\psi)\; \langle  
e^{-\frac{i}{2}J\sum_{a}^{n}(\phi_a-\psi_{a})^2}\rangle \ .
	\non
\]
Note that $g$ measures the deviation from Gaussianity of $\Omega$, 
so that obtaining a non-quadratic form of $g$ means going beyond
the Gaussian approximation of \cite{wuloring} and \cite{strattI}.
The integral self-consistent equation for $g$ is,
\beq
g(\vec\phi)=\int d\vec\psi \;e^{i\frac{\lambda}{2}\sum_a^n\psi_a^2 + 
g(\vec\psi)} \;
\langle  e^{-\frac{i}{2}J\sum_{a}^{n}(\phi_a-\psi_{a})^2} \rangle \ .
	\label{nastro}
\eeq
We assume now that $g$ depends only on the modulus $x$ of 
the replica-vector $\vec\phi$, so that we can perform the angular part 
of the integral, then average over the disorder distribution
$p_\eta (J)$ and finally let $\eta\to 0$, to get:
\beqa
g(x)= \frac{\Gamma(2-c)}{c(c-1)}
\left[ \left(\beta_s+\frac{ix^2}{2}\right)^c - \beta_s^c \right]
\non \\ 
 - x\int_0^\infty dy \; K(x,y) \; e^{i\frac{\lambda}{2}y^2 + g(y)}
	\label{lei}
\eeqa
\beqa
&&\phantom{vadavialku}
K(x,y) = \frac{ \Gamma(2-c) (1+c)}{\Gamma(\frac{3-c}{2})
	\Gamma(\frac{3+c}{2})}	\; \frac{xy}{4} \; \times
\non \\
&&\int_0^1 dt \; \left(\frac{t}{1-t} \right)^\frac{1-c}{2} 
\left\{ \left[ \beta_s+\frac{i}{2}(x^2+y^2)\right]^2 + 
t x^2 y^2 \right\}^\frac{c-2}{2} \ .
	\non
\eeqa
These two equations are the main result of this letter.
It is possible to prove analytically that 
$g(x)\sim -x^{2c}$ for $x\gg 1$, thus proving 
that $\Omega$ is definitely {\it not} a Gaussian distribution. 
Notably, we have been able to numerically 
solve the equation for $g(x)$ without any further approximation. 
Indeed, eq.(\ref{lei}) has the form of a fixed-point
equation, so that it is tempting trying to solve it numerically 
by iteration. This is what we have done, discretizing
the function $g$ and the kernel $K$ on a lattice. We have found 
that the convergence is rather fast and very weakly dependent on
the small imaginary part $\epsilon$ of $\lambda$. Indeed, by setting 
directly $\epsilon=0$, the results are very satisfactory and we are 
able to obtain the solution up to arbitrary precision. 
Once obtained $g$ for a given value of $\lambda$ we have computed 
the spectrum $D(\lambda)$, using the formula,
\[
D(\lambda)=
{\rm Re} \left\{ \frac{1}{\pi}
\int_0^\infty dx\; x\; e^{i\frac{\lambda}{2}x^2 + g(x)} \right\} \ ,
\]
which follows from the definitions of $D$, $\bf G$ and $g$.
The results are shown in Fig.1, where we have plotted the INM spectrum
$D$ as a  function of $\lambda$, for $m=12$.
\begin{figure}
\begin{center}
\leavevmode
\epsfxsize=3in
\epsffile{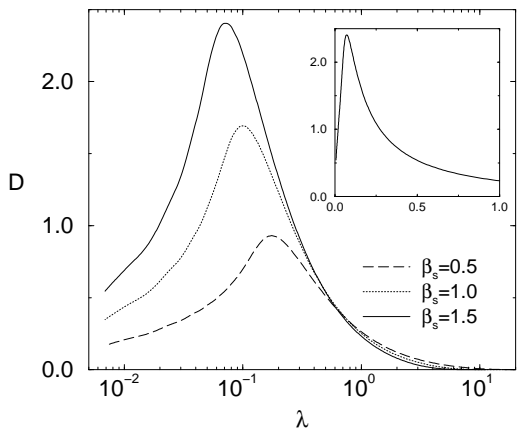}
\caption{The INM spectrum $D$ as a function of $\lambda$ for
different values of the scaled temperature $\beta_s$; $m=12$ and $\epsilon=0$. 
The plot is in log-linear scale. Inset: $D(\lambda)$ for $\beta_s=1.5$ in 
linear-linear scale. The spectrum vanishes at $\lambda=0$.}
\label{fig1}
\end{center}
\end{figure}
The spectrum has positive support because $d=1$ and it depends on the
scaled inverse temperature $\beta_s$ in the expected way: for low
temperatures (high $\beta_s$) the collisions among particles 
are weaker, so that the spectrum is peaked on lower value of the
eigenvalues. On the other hand, the tail for large $\lambda$ is 
larger at higher temperature.
We have found that $D(\lambda)\sim e^{-\alpha \lambda}$ for $\lambda\gg 1$, 
but we have not been able to express $\alpha$ as a function of the parameters
$\beta$ and $m$, even if it is numerically evident that $\alpha$ is a 
monotonically increasing function of $\beta_s$.

A crucial task is now to check whether the result we have 
found is correct. To this aim we have done extensive numerical 
simulations. Once drawn a matrix $\bf J$ with probability (\ref{pj}), 
we build $\bf A$ and diagonalize it numerically. 
Since the spectrum has huge tails for large eigenvalues, it is 
convenient in order to compare simulations with analytic 
results to consider the probability distribution $\pi$ of 
$\mu\equiv\ln \lambda$, that is $\pi(\mu)=D(e^\mu) e^\mu$. 
In Fig.2 we plot $\pi(\mu)$ as obtained from the analytic form of 
$D(\lambda)$, together with the one obtained from numerical simulations.
The two curves are in excellent agreement confirming the validity of 
our result. Besides, we show in the inset of Fig.2 the numerical 
spectrum obtained with the original value of 
$b=m/(m+2)$. 
The result justify the sensibility of the approximation $b\sim 1$.
\begin{figure}
\begin{center}
\leavevmode
\epsfxsize=3in
\epsffile{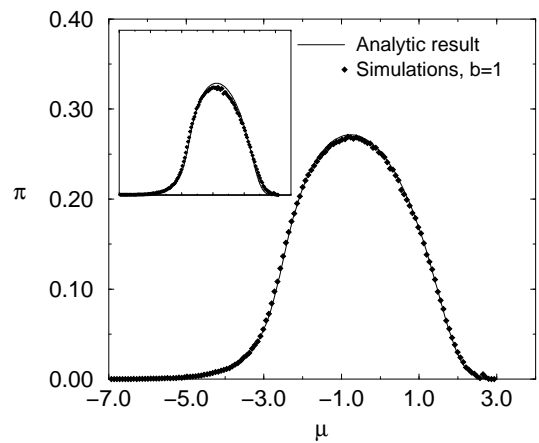}
\caption{Numerical simulations vs. analytic solution. 
We plot here for $b=1$ the probability distribution $\pi(\mu)$, 
with $\mu=\ln \lambda$; $N=600$, $\eta=10^{-4}$, $\beta_s=1$ and $m=12$.
Inset: on the same scale, analytic result for $b=1$ vs. simulations 
performed with $b=m/(m+2)$.}
\label{fig2}
\end{center}
\end{figure}
An equation similar to (\ref{nastro})
have been derived in \cite{brII},
with a different method and within a different context. 
Also in that case the distribution of the disorder was diluted and
translationally invariant, but the explicit probability distribution
$p(J)$ was bimodal,
$p(J)=\delta(J) + \frac{p}{N}[\delta(J-1/p) -\delta(J)]$,
where $p$ was the connectivity \cite{brII}.
Yet, even with this simple bimodal distribution the spectrum had not 
be worked out explicitly until now. 
Remarkably, we have been able to compute exactly the spectrum 
associated to this 
bimodal distribution, by numerically solving the corresponding
self-consistent equation \cite{brII}. 
The spectrum is shown in Fig.3 and can be compared with the
approximated solution of \cite{biroli}. 
Note that in the liquid spectrum there is no trace of the small 
tails oscillations present in the bimodal case \cite{biroli}.
\begin{figure}
\begin{center}
\leavevmode
\epsfxsize=3in
\epsffile{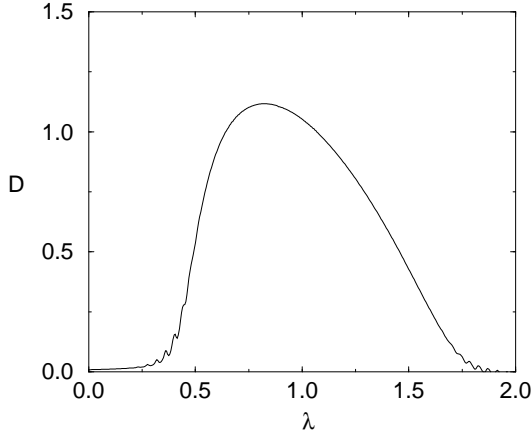}
\caption{Analytic solution. The spectrum in the case of the bimodal 
distribution, with $p=20$ and $\epsilon=0.005$. The small non-zero
value of $D$ for $\lambda\to 0$ is due to the non-zero value of $\epsilon$,
since the limit $\epsilon\to 0$ is very delicate in the bimodal case.}
\label{fig3}
\end{center}
\end{figure}
An important issue is the
analysis of the localization properties of the eigenfunctions. 
With respect to this an important quantity is the average inverse 
participation ratio (IPR) $Y(\lambda)$, which provides information on the 
nature of the eigenfunctions and can be easily computed via numerical 
simulations.
The IPR is defined as, 
$Y(\lambda_\alpha)=\sum_{i=1}^N \left(w_\alpha^i\right)^2$, 
where $\alpha=1\dots N$ is the eigenvalue index and 
$w_\alpha^i=[\langle \lambda_\alpha | i \rangle]^2 $ is the weight of
site $i$ in the eigenfunction $| \lambda_\alpha \rangle$. 
\begin{figure}
\begin{center}
\leavevmode
\epsfxsize=3in
\epsffile{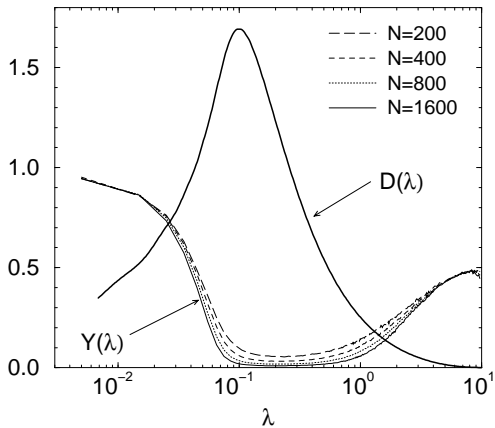}
\caption{The IPR $Y$ as a function of the
eigenvalue $\lambda$ at different values of $N$. $\eta=10^{-4}$, 
$\beta_s=1$ and $m=12$. The thick curve is the corresponding 
spectrum $D(\lambda)$.} 
\label{fig4}
\end{center}
\end{figure}
In Fig.4 we plot $Y$ as a function of $\lambda$. 
It is clear that there are two localizations edges, 
separating a central region of extended eigenvalues from the 
tails where localized states are present. 
For $\lambda\to 0$ the IPR goes to
one and this corresponds to a single particle 
nearly decoupled from the rest of the system. 
Note that the limit $Y\to 1$ for $\lambda\to 0$ is smooth, due to the 
fact that the $J$ distribution we consider is divergent at $J=0$,
so that it never happens that one single element in a row of
$\bf J$ is of order one, with all the others equal to zero. 
This behaviour in the left tail
cannot be present in the case of a bimodal distribution.
On the other hand, the localized states of the right tail correspond
to pairs of very strongly interacting particles and this naturally
leads to an IPR equal to $1/2$.
A more detailed discussion of localization, following
\cite{anderson}, will be presented elsewhere. 

In this letter we have outlined a general method to study analytically 
the INM spectrum of a liquid at low densities. We have succeeded in 
an exact computation in the simple $d=1$ case, but our method is 
suitable to be extended to dimensions larger than one, where the 
Hessian is not positive defined and negative eigenvalues exist.
The presence on negative modes in three-dimensional systems 
is particularly relevant in connection with the glass transition 
\cite{keyes}. Indeed, it has been argued in the context of the INM 
approach, that the Mode Coupling transition, marking the crossover 
from a non-Arrhenius behaviour of the viscosity to an Arrhenius one, 
occurs when the fraction of negative delocalized modes (the only
ones related to collective particles diffusion) drops to zero 
\cite{bembenek-loca}. In view of this, an approach as the one presented 
here, able to investigate also the properties of the localized modes, 
can prove extremely useful.
We will address the extension of our method to higher dimensions 
in a future work.

The work of AC and IG was supported by EPSRC Grant No. GR/K97783.

%----------------------------------------------------------------

%----------------------------------------------------------------

\end{document}